\documentstyle [aps]{revtex}

\newtheorem{defn}{Definition}
\newtheorem{rem}{Remark}
\newtheorem{theo}{Theorem}

\newcommand{\Rn}{{\rm I\!R }} 
\newcommand{\Cn}{{\setbox0=\hbox{
$\displaystyle\rm C$}\hbox{\hbox
to0pt{\kern0.6\wd0\vrule height0.9\ht0\hss}\box0}}} 
\newcommand{\idty}{\hat{\rm 1\mskip-4mu l}} 
\newcommand{\Zn}{{\hbox{$\sf\textstyle Z\kern-0.3em Z$}}} 

\newcommand{\Tr}{{\rm Tr}\,} 

\newcommand{\cA}{{\cal A}}
\newcommand{\cB}{{\cal B}}

\newcommand{\cH}{{\cal H}}
\newcommand{\cK}{{\cal K}}

\newcommand{\cS}{{\cal S}}

\newcommand{\aH}{\cH_{\cA}}
\newcommand{\bH}{\cH_{\cB}}
\newcommand{\abH}{\cH_{\cA + \cB}}
\newcommand{\BHa}{\cB(\aH)}
\newcommand{\BHb}{\cB(\bH)}
\newcommand{\pBH}{{\cB({\cH}_1)}}
\newcommand{\dBH}{\cB(\cH_2)}
\newcommand{\BHab}{\cB(\cH_{\cA + \cB})}
\newcommand{\aS}{\cS_{\cA}}
\newcommand{\bS}{\cS_{\cB}}
\newcommand{\abS}{\cS_{\cA + \cB}}
\newcommand{\abSs}{\abS^{sep}}

\begin{document}
\title{
Some remarks on separability of states\\}
\author{W\l adys\l aw A. Majewski\cite{*}}
\address{Instytut Fizyki Teoretycznej i Astrofizyki\\
Uniwersytet Gda\'nski,
ul. Wita Stwosza 57, 80-952 Gda\'nsk, Poland
}
\maketitle
\begin{abstract}
Several problems concerning separable states are clarified
on the basis of Choi's scheme and old Kadison and Tomiyama results.
Moreover, we generalize Terhal's construction of positive maps.
\end{abstract} 

\section{Introduction}
\label{I}
In the analysis
of quantum mechanical properties of composite systems there is a strong demand
for a characterization of various subsets of states. Among properties of states,
perhaps the most intriguing phenomenon is that of entanglement.
We recall that, in physical terms, this notion 
reflects inability of local preparation of the state.
Probably, the most famous example of states with entanglement is that with violation
of Bell inequalities (cf. \cite{Peres1}). Other nice applications can be found in the theory
of decoherence or in quantum information theory.
This makes clear why one of the problems in foundations of quantum theory is 
a classification and characterization of the entangled quantum states. In this context 
it has been shown (cf. \cite{Peres2}, \cite{HHH}, \cite{DVSS}) 
that there should be a strong connection between 
classification of entanglement of quantum states and the structure of positive maps.
The problem is complicated since it is necessary to take into account 
some basic facts concerning
specific properties of tensor product of algebras.
Unfortunately, such approach involves mathematically advanced concepts and therefore
it is not easily accessible to most physicists working on applications
of Quantum Mechanics to Quantum Information or Quantum Optics.
Moreover, the theory of positive linear maps of $C^*$ algebras appears in the physical
literature in a rather scattered and usually special form, the best known examples
are states, dynamical maps, and $^*$-representations. 

The purpose of this paper is to present, in plain words, state-of-the-art of the 
relevant parts of theory of positive maps as well as to clarify the present day 
characterization of entanglement. The structure of the paper is the following:
in Section II we review some results in the theory of positive maps
and show how these facts can be applied to a characterization of some singled out
subsets of states. Section III is devoted to indicate that the family of separable states
is really very small. Finally, in Section IV, 
some remarks concerning recent results on positive maps will be given.
Moreover, we argue that the problems associated to the separability follow directly
from the rule saying that a composite system is described by tensor product.

Finally, we want to emphasize that we address this paper to physicists
working on application of Quantum Mechanics. Therefore, the presented material does not 
include neither general results nor can be considered as a full
account of all mathematical aspects of the theory. However, all results are rigorous.  
 
\section{Application of positive maps to the description of separability }
\label{II}

We consider a composite system $\cA + \cB$ consisting of
two subsystems $\cA$ and $\cB$ respectively.
The algebra of all observables is given by $\BHab$, $\BHa$, and $\BHb$,
for $\cA + \cB$, $\cA$, and $\cB$ respectively where $\cB(\cH)$
stands for the set of all linear bounded operators on a Hilbert space $\cH$.
Moreover, $\abH$ will be identified with $\aH \otimes \bH$.
We note that the particular case $dim \aH =n < \infty$,
$dim \bH = m < \infty$ (dim stands for dimension) is the standard
assumption in an analysis of entanglement states in Quantum Information.
Recall that a state (see \cite{buba}) of a system, say $\cA$, 
is defined as a linear, complex-valued map
$\phi : \BHa \to \Cn$ such that $\phi(A^*A) \ge 0$, 
$A \in \BHa$, and $\phi(\idty) = 1$
($\idty$ is the identity operator in $\cB(\cH)$).
Usually, in physical literature, a state $\phi$ is assumed to be of 
the form $\phi(A) = \Tr \varrho A$ where $\Tr$ stands for the trace, 
$\varrho$ for a density matrix. This fact is used frequently as a legitimacy for 
identification of a state with a density matrix.
We denote the family of all states of
$\cA$ ( $\cB$, $\cA + \cB$) by $\aS$ ($\bS$, $\abS$ respectively).
Finally, we will need two definitions of positivity.
Let $\Psi: \pBH \to \dBH$ be a linear map ($\cH_i$ is a Hilbert space).
$\Psi$ is positive when $\Psi: \pBH^+ \to \dBH^+$ where $\cB(\cH_i)^+$
denotes all positive operators in $\cB(\cH_i)$, $i=1,2$. 
Now let $id_k$ be the identity map on $M_k(\Cn)$ (set of all $k \times k$
matrices). We define the map $id_k \otimes \Psi : M_k(\Cn) \otimes \cB(\cH) \to
M_k(\Cn) \otimes \cB(\cH)$, for $k=1,2,..$ by
\begin{equation}
(id_k \otimes \Psi)(\sum_i \sigma_i \otimes A_i) = \sum_i
\sigma_i \otimes \Psi(A_i),
\end{equation}
where $\sigma_i \in M_k(\Cn)$ and $A_i \in \cB(\cH)$.
The map $\Psi$ is $k$-positive when $id_k \otimes \Psi$ is positive.
The map $\Psi$ is completely positive when $\Psi$ is $k$-positive for all $k=1,2,...$
\vskip 0.5cm
Having the defined notion of state we can ask how $\abS$ is related to
$\aS$ and $\bS$. This is rather deep question and we start its analysis
with the following simple exercise (cf. \cite{KR} , 11.5.11):
let $dim \aH = 2 = dim \bH$. Define the state $\varphi$ in $\abS$
as $\varphi(A \otimes B) = {1\over 2}\bigl( (A \otimes B)(e_1 \otimes f_1
+ e_2 \otimes f_2), e_1 \otimes f_1 + e_2 \otimes f_2\bigr)$
where $\{e_i\}_{i=1}^2$ is a basis in $\aH$,
$\{ f_i\}_{i=1}^2$ a basis in $\bH$.
In other words, $\varphi(\cdot)$ is a state determined by the
vector ${1 \over {\sqrt 2}}(e_1 \otimes f_1 + e_2 \otimes f_2) \in
\abH$. We denote by $\varphi_0$ a convex combination
of product states, i.e.
\begin{equation}
\varphi_0 = \sum a_n \omega_n^{\cA} \otimes \omega_n^{\cB}
\end{equation}
where $\omega_n^{\cA}(A) = (Ax_n, x_n)$ for $A \in \BHa$, $x_n \in \aH$
and analogously for $\omega_n^{\cB}$. $\{a_n\}$ are non-negative numbers
such that $\sum_n a_n = 1$. Then one can show (see (\cite{KR}) that
\begin{equation}
|| \varphi - \varphi_0|| \ge {1 \over 4}.
\end{equation}
This clearly shows that the state $\varphi$ of
$\BHa \otimes \BHb$
lies at norm distance at least ${1 \over 4}$ from the convex hull
of the set of product states.
Consequently, the family of convex combinations of product states 
is a proper subset of the set of all states on $\BHa \otimes \BHb$
and moreover this family is not a norm dense subset in $\abS$, i.e. , in general,
arbitrary state can not be approximated in norm by convex combination
of product states.
This result shows the crucial property of the space of states of the 
composite system. Therefore we define
\begin{defn}
A state $\varphi$ defined on $\BHa \otimes \BHb$ is said to be separable 
if it can be written as
\begin{equation}
\label{cztery}
\varphi = \sum_n a_n \varphi_n = \sum_n
a_n \varphi^{\cA}_n \otimes \varphi^{\cB}_n
\end{equation}
where $\varphi_n = \varphi_n^{\cA} \otimes \varphi_n^{\cB}$ with
$\varphi_n^{\cA}$ ($\varphi^{\cB}_n$) a state on $\BHa$ ($\BHb$),
and $a_n$ are positive numbers that sum to one. The set of all separable states
, for the system $\cA + \cB$ will be denoted by $\abS^{sep}$.
A state which can not be written as in (\ref{cztery}) is called entangled or 
equivalently non-separable.
\end{defn}
 
Peres (\cite{Peres2}) showed that a necessary condition for a density matrix of a 
bipartite system to be separable is for its partial
transpose to be a density matrix where the partial transpose
is just the transposition of the matrix representing the state 
of either subsystem $\cA$ or subsystem $\cB$.
For $2\otimes 2$ and $2\otimes 3$ systems this condition is also 
sufficient (\cite{HHH}). This is the point where a part of theory of 
positive maps enters to the problem.
To make this point more clear we restrict ourselves, for a moment,
to finite dimensional case (so $dim\aH = n < \infty$, $\BHa$ can 
be taken as $M_n(\Cn) \equiv M_n$ and the same for subsystem $\cB$)
and we recall the following scheme which is taken from Choi's lecture (\cite{Choi}):
\vskip 0.5cm

\hyphenation{assuming}
\hyphenation{functionals}
\hyphenation{positive}
\begin{tabular}{p{3cm}cccp{3.7cm}}
$L(M_n,M_k)$ &
$\longleftrightarrow$ &
$M_n\otimes M_k$ &
$\longleftrightarrow$ &
linear functionals on $M_n\otimes M_k$ 
\\[1mm]
 $\quad$   $\bigcup$ & &&& $\quad$ $\bigcup$
\\[1mm]
hermitian preserving linear maps
& $\longleftrightarrow$ &
$(M_n\otimes M_k)^{\rm h}$ &
$\longleftrightarrow$ &
linear functionals assuming real values on $(M_n\otimes M_k)^{\rm h}$ 
\\[1mm]
$\quad$ $\bigcup$ &&&& $\quad$ $\bigcup$
\\[1mm]
positive linear maps &
$\longleftrightarrow$ &
? &
$\longleftrightarrow$ &
linear functionals assuming positive values on $M_n^+\otimes M_k^+$
\\[1mm]
$\quad$ $\bigcup$ &&&& $\quad$ $\bigcup$
\\[1mm]
completely positive linear maps &
$\longleftrightarrow$ &
$(M_n\otimes M_k)^+$ &
$\longleftrightarrow$ &
linear functionals assuming positive values on $(M_n\otimes M_k)^+$
\end{tabular}

\bigskip
where $\longleftrightarrow$ denotes a $1-1$ correspondence,
$\bigcup$ is an inclusion, $M_n^h$ stands for all hermitian matrices
in $M_n$, $M_n^+ \otimes M_k^+ = \{ \sum A_i \otimes B_i,\quad A_i \in M_n^+,
B_i^+ \in M_k^+ \}$,
while $L(M_n,M_k)$ stands for linear maps from $M_n$ to $M_k$.

\vskip 0.5cm

\begin{rem}
Even this finite dimensional case clearly shows strong relations among:
states, some singled out subsets of states, positive maps, and 
finally completely positive maps. This can be taken as a hint that 
the core of the considered problem is related to some special
properties of the order (defined by positive elements) in tensor 
products. Moreover, one can expect that definition of order
in tensor products leads to some serious problems.
This is so, and for a general review of that question
we refer Wittstock's lecture in (\cite{Witt}). Later on, in the next Section,
we come back to that point.
\end{rem}
\vskip 0.3cm
Choi's scheme gives the following simple fact relating separable states 
with the order in tensor product. Let $\varphi$ be in $\abSs$,
then (we recall that $M_n^+ \otimes M_k^+ \subset
(M_n \otimes M_k)^+$ where $\subset$ is the proper inclusion)
\begin{equation}
A \in M_n^+ \otimes M_k^+ \quad implies \quad \varphi(A) \ge 0.
\end{equation}
Also, 
\begin{equation}
\varphi \circ (\Psi_n \otimes \Psi_k)|_{M_n^+ \otimes M_k^+} \ge 0
\end{equation}
for $\Psi_n$ ($\Psi_k$) positive maps on $M_n(\Cn)$ 
($M_k(\Cn)$ respectively).
Obviously, the same holds if one replaces $M_n$ by $\BHa$ 
with not necessary $dim\aH < \infty$, etc.

\bigskip
Thus Choi's scheme leads to the origin of Peres-Horodeckis characterization
of separable states. Namely, let us recall that the transposition map
$\tau: a_{ik} \mapsto \tau(a_{ik})= a_{ki}$
is a positive map but not even two-positive, so not completely
positive (cf. \cite{Choi2}). Take a state $\varphi \in \abS$ and $\psi \in \abSs$.
Denote by $\tau^{\cB}$ the transposition map acting
on the set of observables associated with the subsystem $\cB$.
Then, according to Choi's scheme $\varphi \circ id^{\cA} 
\otimes \tau^{\cB}$ is a linear functional taking positive values on 
$M_n^+ \otimes M_k^+$ but not in general on $(M_n \otimes M_k)^+$.
Therefore, separable states can differentiate positivity from 
completely positivity.
This idea was used in

\begin{theo}{(Horodecki's. \cite{HHH})  }
A density matrix $\varrho$ on $\aH \otimes \bH$ is entangled iff there 
exists a positive linear map $\cS : \aH \to \bH$ such that
\begin{equation}
(id_{\cA} \otimes \cS)\varrho
\end{equation}
is not positive semidefinite. Here $id_{\cA}$ denotes the identity map on $\BHa$.
\end{theo}

\vskip 0.5cm

Let us turn again to Choi's scheme. To complete partly the scheme we recall that 
one can endow the set of all matrices, say $n \times n$, with the so called 
Hadamard product (cf. \cite{MM}). Let $A,B \in M_n(\Cn)$, then the matrix
$H = A \ast B \in M_n(\Cn)$ is called Hadamard product of $A$ and $B$ if its $(i,j)$ 
entry is equal to $a_{ij}b_{ij}$. Let $\Psi: M_n \to M_n$ be a positive map, then
$A \ast \Psi(A)$ is also positive for all $A \in M_n^+$.
Conversely, for a fixed $A \in M_n^+$ the map $S_A : M_n \to M_n$
defined by
\begin{equation}
S_A(B) = A \ast B.
\end{equation} 
leads (as it will be shown) to a positive map.
Thus we get a nice correspondence:
\begin{center}
positive maps in $L(M_n,M_n)$  $\leftrightarrow$ $M_n^+ \ast M_n^+$.
\end{center}
To see the positivity of $S_A$ and to
understand this result in the context of separability problem let us reconsider 
another Kadison-Ringrose exercise (cf. \cite{KR} 2.8.41). Namely, let $\cH$ be a Hilbert space
(not necessary of finite dimension) with an orthonormal basis $\{e_1,e_2,...\}$.
We denote by $[a_{ij}]$ and $[b_{ij}]$ the matrices of $A,B \in \cB(\cH)$ with respect to
the basis $\{e_i\}$. Next, let us consider the orthonormal system
$\{e_i \otimes e_i \}_i $ in $\cH \otimes \cH$ and we denote by $\cK$ the
subspace of $\cH \otimes \cH$ spanned by the system $\{e_i \otimes e_i \}$. Finally,
the projection from $\cH \otimes \cH$ onto $\cK$ will be denoted by $P$. Then solving
the Kadison-Ringrose exercise 2.8.41 we find that the matrix elements
$\{ t_{ij} \}$ of the operator $(P(A \otimes B)P|_{\cK}) \equiv T$ with respect to the basis 
$\{ e_i \otimes e_i \}$ of $\cK$ are
\begin{equation}
t_{ij} = a_{ij}b_{ij} \equiv([a_{ij}] \ast [b_{ij}])_{ij}.
\end{equation}
Clearly, $\{t_{ij} \} \ge 0$ provided that $\{ a_{ij} \} \ge 0$ and
$\{ b_{ij} \} \ge 0$.

Now, let $A$ be a density matrix $\varrho$, $B$ a density matrix $\sigma$, both
acting on a Hilbert space $\cH$. Then, dropping for simplicity the normalizating
factor, we got a hint that there are density (nonseparable) matrices satisfying 
Peres condition (note that $P$ is not in $\cB(\cH)^+ \otimes \cB(\cH)^+$; cf.
\cite{KR}, exercise 11.5.10).

\section{How generic separable states are?}
\label{III}

Among questions concerning the separability, great effort has been done to determine
degree of largeness of the family of separable states in the set of 
all states (cf. \cite{ZHSL}, \cite{CH}). 
We have seen, in Section II, that $\cS^{sep}$ is not a norm-dense subset
of $\cS$. However, it appear highly desirable to study this question
with a concept of convergence which would be more accessible to experimental
verifications. Namely, from physical point of view, it would be reasonable to say that
 a sequence of separable states $\{ \varphi_n \}$ approximates a state $\varphi$
if, given any observable $A$ in $\cB(\cH)$ and any $\epsilon >0$ one can find
an integer $N(A,\epsilon)$ such that
\begin{equation}
|\varphi_n(A) - \varphi(A)| < \epsilon \quad for \quad all \quad n \ge N(A, \epsilon).
\end{equation}
In mathematical terms, it would mean: is the set $\cS^{sep}$ weakly $^*$-dense in $\cS$?
In Section II it was shown that a description of $\cS^{sep}$ (as well as of $\cS$) is 
closely related to distinguished subsets of positive observables in the tensor
product $\BHa \otimes \BHb$. So to solve the just posed question we should have
a characterization of positiveness in the tensor products.
The desired result follows from old Tomiyama theorem (see \cite{To}). To quote his result,
unfortunately, some additional definitions are necessary.
We define (cf. \cite{KR}) n-state of $\cB(\cH)$ to be a matrix $[\vartheta_{ij}]$
of linear functionals on $\cB(\cH)$ such that the matrix $[ \vartheta_{ij}(A_{ij}) ]$
is semi-positive defined when $[ A_{ij} ] \in M_n(\cB(\cH))^+$
(so the matrix $[A_{ij} ]$ having entries in $\cB(\cH)$, is also nonnegative) and
$\vartheta_{ii}(\idty) =1$ for $i=1,2,..$. If the normalization is omitted we get the notion
of n-positive functional.

To get an example of such $n$-state let us pick $\{x_1,x_2,... \}$ 
a set of unit vectors in $\cH$. Denote by $\omega_{x_i,x_j}$ the following functional
$\omega_{x_i,x_j}(A) = (Ax_i,x_j)$, $A \in \cB(\cH)$.
Then, one can easily check that $\{ \omega_{x_i,x_j}\}$ is an
$n$-state on $\cB(\cH)$.
\smallskip
\begin{rem}
\label{uwaga}
We have seen, in Section II, the great significance of complete positive
maps in the analysis of separable states. In that context the following
observation seems to be important: a linear map $\Phi : \cB(\cH_1)
\to \cB(\cH_2)$, such that $\Phi(\idty) = \idty$,
is completely positive if and only if $\{ \vartheta_{ik} \circ \Phi \}$
is n-state for each n-state $\varrho_{ik}$ of $\cB(\cH_2)$.
\end{rem}
\smallskip
Let $\varphi$ be a functional on $\cB(\cH_1) \otimes \cB(\cH_2)$.
$\varphi$ is called a positive functional of order $n$ if $\varphi$ is expressed
as
\begin{equation}
\varphi = \sum_{i,j=1}^n \phi_{ij} \otimes \psi_{ij}
\end{equation}
where $\{ \phi_{ij}\}$ and $\{\psi_{ij} \}$ are $n$-positive functionals
on $\cB(\cH_1)$ and $\cB(\cH_2)$ respectively.

Now, in this setting, the Tomiyama theorem asserts that the order in
$\cB(\cH_1) \otimes \cB(\cH_2)$ (so the distinguished family
of positive observables of the composite system) is determined by the set of
positive functionals of order n where n is equal to $min\{ dim \cH_1, dim\cH_2 \}$.

Now to get the promised estimation of degree of largeness of $\cS^{sep}$
 we have to recall another
Kadison's result. Namely, a family of states which determines the order 
is weak $^*$ dense and conversely a weak $^*$ dense family of states determines the 
order in the set of observables.
Combining Kadison and Tomiyama results one can see that $\cS^{sep}$ can  
not be weakly dense subset in $\cS$ unless one of the subsystems is one dimensional
(so classical one, cf. also \cite{Maj}).

To make more clear the above conclusion let us consider the favorite model in Quantum
Information. We assume that, say, $dim \cH_1 = 2$, $dim \cH_2$ can be arbitrary.
Then the order of $\cB(\cH_1) \otimes \cB(\cH_2)$ is determined by functionals of the form
\begin{equation}
\varphi = \sum_{i,j=1}^2 \phi_{i,j} \otimes \psi_{ij}.
\end{equation}
Take for $\{ \phi_{ij} \}$ the 2-state with $\phi_{ij}$ being the same state
$\eta_1$ of $\cB(\cH_1)$ and analogously for $\{ \psi_{ij} \}$.
Then, it is clear that the family $\cS_0$ determining 
the order in the tensor product contains as
 a proper subset the family of separable states, $\cS^{sep} \subset \cS_0$ but not
$\cS^{sep} \ne \cS_0$. Therefore, in general, one can not
approximate a state $\varphi$ by separable states in a way which is accessible to 
experimental verifications. By the way, the same argument leads to similar conclusion for 
the corresponding property of entangled states.

There is a remarkable relation between $k$-functionals and a certain class
of states with Schmidt rank k which were considered recently (see for example
\cite{DVSS}).
Namely, let $|\psi> = \sum_{i=1}^k \sqrt{\lambda_i} |x_i> \otimes |y_i>$
where $\{ |x_i> \}$ and $\{ |y_i> \}$
are orthonormal systems. We observe
\begin{eqnarray}
Tr(|\psi><\psi| \cdot a\otimes b) = <\psi|a \otimes b |\psi>
= \sum_{ij} \sqrt{\lambda_i \lambda_j}\omega_{x_i,x_j}(a) \omega_{y_i,y_j}(b) \nonumber \\
=\sum \sqrt{\lambda_i \lambda_j}(\omega_{x_i,x_j} \otimes \omega_{y_i,y_j})(a \otimes b).
\end{eqnarray}
Clearly, $Tr \{ |\psi><\psi| \cdot \}$ is a positive functional of order $k$
while $\{\omega_{x_i,x_j} \}$ and $\{ \omega_{y_i,y_j} \}$
are $k$-states.
Let $\Phi$ be a positive normalized map and $\omega_{x_i,x_j}$ a $k$-state. 
Then it is well known that
$\Phi$ is $k$-positive iff $\omega_{x_i,x_j} \circ \Phi$ is again $k$-state 
(cf. Remark \ref{uwaga}).
This implies that 
\begin{equation}
Tr(|\psi><\psi| a\otimes \Phi(b)
\end{equation}
is another positive functional of order $k$ if $\Phi$ is $k$-positive.
Consequently, this remark and Kadison and Tomiyama results
provides the general context of the recent characterization of
$k$-positivity given in (\cite{DVSS}).

\vskip 1cm
To get another approximation
procedure as well as 
another characterization of separable states one can proceed as follows.
Let $\varrho$ be a density matrix on $\cH_1 \otimes \cH_2$ and let $\{ e_i \}_i $
be a basis in $\cH_2$. Define a linear isometry $U_i : \cH_1 \to \cH_1 \otimes \cH_2$
by $U_ix = x \otimes e_i \in \cH^i$ 
for $x \in \cH_1$ where $\{ \cH^i \}$
is a family of pairwise orthogonal subspaces in $\cH_1 \otimes \cH_2$.
Then $U_i^*$ is a linear map of $\cH_1 \otimes \cH_2$ in $\cH_1$ such that
$U^*_k(\cH_1 \otimes \cH_2 \ominus \cH^l) = 0$ for $k \ne l$.
For $\varrho \in \cB(\cH_1 \otimes \cH_2)$ we define
\begin{equation}
\label{dwanascie}
\widehat{\varrho_{ik}} = U^*_i \varrho U_k \in \cB(\cH_1).
\end{equation}
In other words, the density matrix $\varrho$ can be represented as a semipositive
matrix $\{ \widehat{\varrho_{ik}} \}$ with operator-valued entries.

Assume $\varrho$ to be of the form $\varrho= \varrho_1 \otimes \varrho_2$ with
$\varrho_i \in \cB(\cH_i)$, $i = 1,2.$ The the corresponding
matrix has the very special form:
\begin{equation}
[ \widehat{\varrho_{ik}}] = [ \lambda_{ij} \varrho_1 ]
\end{equation}
where $[\lambda_{ik} ]\equiv [(e_i,\varrho_2 e_k)]$ 
is semipositive defined matrix with $\Cn$-valued entries,
$\varrho_1$ is a density matrix in $\cB(\cH_1)$.
Therefore, such matrix has commutative entries.
Furthermore, any separable density matrix is just a convex combination of matrices
of such special form.

Now we in position to study another approximation procedure. Assume $\varrho$ can be 
approximated by separable density matrices. It would mean that, in just described
matrix representation, one can find a family of semipositive defined matrices 
$[ \lambda_{ik}^{\alpha} ]$ with
$\Cn$-valued entries, a family of density matrices $\{ \varrho^{\alpha} \}$
(on the Hilbert space associated with the arbitrary but fixed subsystem) such that
\begin{equation}
\widehat{\varrho_{ik}} - \sum_{\alpha} \lambda_{ik}^{\alpha} \varrho^{\alpha}_1
\end{equation}
is small. Clearly, for high dimension this is a hopeless task.
For low dimensions, for some subsets of states and (or) for some finite
accuracy one can get such approximation.

\section{General positive maps and concluding remarks}
\label{IV}
In previous Sections, general positive maps are frequently used for an analysis
of separable and non-separable states. On the other hand,
completely positive maps are admittedly the "physical" ones (cf. {\cite{Lin}).
So one can suspect us for using "non-physical"
tools for a description of some singled out subsets of "physical" states.
To argue that this is not the case we again use another Kadison result (cf. \cite{Kad}).
Namely, let us restrict ourselves to Hamiltonian flows and consider equivalence between 
Schr\"odinger and Heisenberg picture.  Physical maps sending states into states can be taken
to be affine (it takes into account the superposition principle) bijections 
(this property reflects the reversibility of Hamiltonian dynamics).
Their counterparts in Heisenberg picture are linear maps, preserving hermitian
conjugation and anticommutators (in mathematical terms they are
Jordan morphisms). But any Jordan morphism
can be splitted into
$^*$-homomorphism (so a completely positive map) and $^*$-anti-homomorphism
(only positive, not even two positive map, e.g. the transposition 
$\tau$ used in Section II
is an example of $^*$anti-homomorphism).
So this result shows that there is a room for physical maps which are not completetely
positive. 

Having a motivation for positiveness we want to present a general construction of 
a very general linear, positive map $S : \BHa \to \BHb$ (cf. \cite{Ter} and  \cite{Tato}).
For simplicity we assume that $\aH$ and $\bH$ are finite dimensional Hilbert spaces.
However, nearly all results can be straightforwardly generalized to 
infinite dimensional case.
Let us take a hermitian operator $H \in \cB(\aH \otimes \bH)$ such that
\begin{equation}
\label{duality}
H \in \{ A \in \cB(\aH \otimes \bH);\quad \phi(A)\ge 0 \quad for \quad all \quad
\phi \in \abSs \} \equiv \cS_{sep}^d.
\end{equation}
and we assume that $H \not\in \cB(\aH \otimes \bH)^+$.
This is always possible since $\BHa^+ \otimes \BHb^+$  ($\abSs$)
is a proper subset of $\cB(\aH \otimes \bH)^+$ ($\abS$ respectively).
For some further explanation an an example of such a choice
see (\cite{Maj}, another example can be found in (\cite{Ter}). In fact, here,
we are partly following Terhal and Takasaki-Tomiyama idea (\cite{Ter}, \cite{Tato}).

Let $\{ e_i \}$ be an orthonormal basis in $\aH$, $z,x$ arbitrary
vectors in $\aH$, $v,y$ arbitrary vectors in $\bH$.
Let us observe (we take the scalar product to be linear in the second factor)
\begin{eqnarray}
\label{tita}
(z \otimes v, H x \otimes y) = \sum_{ij}\overline{(e_i,z)}
(e_j,x)(e_i \otimes v, H e_j \otimes y) 
= \sum_{ij}(z,E_{ij}x)(v,V^*_i H V_j y) \nonumber\\
=(z\otimes v, \bigl( \sum_{ij} E_{ij} \otimes V^*_i H V_j \bigr) x \otimes y)
\end{eqnarray}
where $E_{ij} \equiv |e_i><e_j|$, $V_i y = e_i \otimes y$ (so $V_i$ is
defined analogously to $U_i$, see definition given prior to (\ref{dwanascie})).
Define the map $S$
\begin{equation}
\label{odwzorowanie}
S(E_{ij}) = V^*_i H V_j.
\end{equation}
Let us note
\begin{equation}
S(E_{ij})^* =(V_i^* H V_j)^* = V_j^* H V_i = S(E_{ji}) = S(E_{ij}^*).
\end{equation}
Therefore to prove positivity of the map $S$ it is enough to show that $S$ maps any 
projector $|f><f|$ ($f \in \aH$) into a positive operator. To this end we note
\begin{eqnarray}
(y, S\bigl(|f><f|\bigr) y) = (y, \sum_{ij} \overline{\lambda_i} \lambda_j
S(E_{ij})y)
= \sum_{ij} \overline{\lambda_i}\lambda_j (y,V^*_i H V_j y) \nonumber\\
\sum_{ij}(\lambda_i e_i \otimes y, H \lambda_j e_j \otimes y)
= (f \otimes y, H f \otimes y) \ge 0
\end{eqnarray}
where the last inequality follows from ({\ref{duality}). Consequently, $S$ is a positive map.
Further we note that (\ref{tita}) implies
\begin{equation}
H=\sum_{ij} E_{ij} \otimes S(E_{ij})
= \sum_{ij}(\idty_{\cA} \otimes S)(E_{ij} \otimes E_{ij}).
\end{equation}
Let us assume additionally that $\aH \subseteq \bH$. Then
\begin{eqnarray}
Tr_{\aH \otimes \bH} (\mu H )= \sum_{ij} Tr_{\aH \otimes \bH}\bigl( \mu
(\idty_{\cA} \otimes S)(E_{ij} \otimes E_{ij})\bigr) \nonumber\\
=\sum_{ij}Tr_{\aH \otimes \bH} \bigl((\idty_{\cA} \otimes S^*)(\mu) 
\cdot E_{ij} \otimes E_{ij}\bigr)
= Tr_{\aH \otimes \bH}\bigl((\idty_{\cA} \otimes S^*)(\mu) 
\cdot P_{\sum_i e_i \otimes e_i}\bigr)
\end{eqnarray}
where $P_{\sum_i e_i \otimes e_i}$ is the projector onto $\sum_i e_i \otimes e_i$.
Since $H \not\in \cB(\aH \otimes \bH)^+$ then one can find a density matrix, say,
$\mu$ such that $Tr_{\aH \otimes \bH} ( \mu H) < 0$. Thus
\begin{equation}
\bigl( \sum_i e_i \otimes e_i, (\idty_{\cA} \otimes S^*)(\mu) \sum_i
e_i \otimes e_i \bigr) <0.
\end{equation}
Consequently, $S^*$ {\it so the map $S$} is not
a completely positive map. We want to emphasize that {\it this result
stems from the fact that $\cS^{sep}$ is a proper subset of the set of all states
$\cS$}; see a remark prior to definition of separable states.
\smallskip
Now we want to show that the above construction provides positive
maps which are not decomposable ones. We recall that a linear positive
map $S$ is decomposable if it can be written as
\begin{equation}
S = S_1 + S_2 \circ \tau
\end{equation}
where $S_1,S_2$ are completely positive maps while $\tau$ is a transposition.
To prove the claim let us take $H\in \cS_{sep}^d$, $H^* = H$
 and a state $\varrho \in \cS$ such that
\begin{equation}
TrH \varrho <0 
\end{equation}
and $(\idty \otimes \tau) \varrho $ is semipositive defined operator.
Here we assume that the Hilbert spaces are of dimension larger than 2.
Namely, it can be shown that for the case $dim \cH_1 = 2$
and $dim \cH_2 = n$, $n \ge 2$,  it is impossible to find density matrix
$\varrho$ such that the assumed conditions are satisfied.
For Hilbert spaces of higher dimension
such a choice is possible, for an example see (\cite{Ter}).

Now we want to argue that the condition:
for a fixed state $\varrho$
\begin{equation}
\label{raz}
Tr\bigl( (\idty \otimes \tau) \varrho)B \bigr) \ge 0
\quad {for \quad any} \quad B \in \cB(\aH \otimes \bH)^+
\end{equation}
implies
\begin{equation}
\label{dwa}
Tr\bigl( (\idty \otimes \tau \cdot T)( \varrho) B \bigr) \ge 0
\quad { for \quad any \quad completely \quad positive \quad map} \quad T
\quad {and} \quad B \in \cB(\aH \otimes \bH)^+.
\end{equation}
To show this we observe
$B \in \cB(\aH \otimes \bH)^+$ if and only if $B = C^*C$ with
$C \in \cB(\aH \otimes \bH)$. So $C = \sum_i E_i \otimes F_i$
and $B= \sum_{ij} E^*_i E_j \otimes F_i^* F_j$.
But (\ref{raz}) implies
\begin{equation}
\label{trzy}
\sum_{ij} Tr \bigl( \varrho (\idty \otimes \tau^*)(E^*_iE_j \otimes F_i^* F_j) \bigr) \ge 0.
\end{equation}
Thus
\begin{equation}
\label{czter}
\sum_{ij} Tr \bigl( \varrho E^*_iE_j \otimes \tau^*(F_i^* F_j) \bigr) \ge 0.
\end{equation}
On the other hand (\ref{dwa}) can be rewritten as
\begin{eqnarray}
\label{piec}
\sum_{ij} Tr\bigl( (\idty \otimes T)(\varrho) \cdot E_i^*E_j \otimes
\tau^*(F^*_i F_j \bigr)
= \sum_m \sum_{ij} Tr \bigl( \varrho \cdot
E_i^* E_j \otimes V_m \tau^*(F^*_i F_j) V_m^* \bigr) \nonumber\\
= \sum_m \sum_{ij} Tr \bigl( \varrho \cdot E^*_i E_j \otimes \tau^*( 
\tau^*(V_m) F^*_i F_j \tau^*(V_m) ) \bigr)
\end{eqnarray} 
where we used the fact that  the transposition $\tau$ is antihomomorphism
and the fact that any completely positive map $T(A)$ can be written
as $\sum_m V^*_m A V_m$ (cf. \cite{Choi3}).
But (\ref{czter}) implies 
\begin{equation}
\sum_{ij} Tr \bigl( \varrho \cdot E^*_i E_j \otimes \tau^*(V^*_m F_i^* F_j V_m) \bigr)
=\sum_{ij} Tr\bigl( \varrho \cdot E^*_i E_j \otimes \tau^* ( G_i^* G_j) \bigr) \ge 0
\end{equation}
where we put $G_i \equiv F_i \tau^*(V_m)$ for any $i$, and this 
completes the proof of our statement.

Let us turn to the question of nondecomposable maps. We assume that 
$(\idty \otimes \tau)\varrho \ge 0$ and the map $S$
defined in (\ref{odwzorowanie}) is decomposable. So $S = S_1 + S_2 \cdot \tau$ 
where $S_1, S_2$ are completely positive maps and $\tau$, as usually, is the transposition.
We observe
\begin{equation}
0> Tr H \varrho = <\sum_i e_i \otimes e_i| \idty \otimes
(S^*_1 + \tau^* S_2)(\varrho) | \sum_i e_i \otimes e_i>.
\end{equation}
But this is a contradiction since the just proved statement implies
\begin{equation}
(S^*_1 + \tau^* \cdot S^*_2)(\varrho) \ge 0
\end{equation}
Consequently, $S$ can not be a decomposable map.

To close our remarks on positive maps let us mention that there is
a nice relationship between such approach to a construction of a class of positive
maps and unextedible systems (cf. \cite{DVSS}, \cite{Ter}). To do this 
we have to recall some Woronowicz results.
Namely, as the general theory of positive maps is very complicated (cf. Choi's scheme), 
the investigation of positive maps based on analysis of extreme positive maps did not 
provide satisfactory results (cf. \cite{St}), Woronowicz (\cite{Wor}) proposed to study
nonextendible positive maps - a map $\Phi$ is nonextendible if it can not be 
written in the form 
\begin{equation}
\Phi(a) = P\tilde{\Phi}(a) P
\end{equation}
where $\tilde{\Phi}$ is another positive map, $P$ is a projection, in a 
nontrivial way (see \cite{Wor} for a precise definition). 
Now assume $dim \aH = n < \infty$,
$dim \bH = n < \infty$. Then the map $S$ defined with $H = \sum_{i=1}^n
|\alpha_i><\alpha_i| \otimes |\beta_i><\beta_i|$ 
for a properly chosen orthonormal systems leads
to a nonextendible map. To prove this statement it is enough to
apply Theorem 3.3 in (\cite{Wor}). To appreciate the notion of nonextendible
positive maps we recall some Woronowicz results: nonextendible
normalized positive map is extreme in the cone of all positive
maps, and any normalized positive map admits a nonextendible extension.
Moreover any Jordan map (normalized positive map $\Phi$
such that $\Phi(a^2) = (\Phi(a))^2$) is nonextendible.

\vskip 1cm
To end this paper, we want to stress that nearly all given results follow, more or less,
directly from mathematical features of tensor products. 
But using the Quantum Mechanics rule
saying that composite systems are described by tensor products we should 
expect that some properties of tensor product have a great impact on a
description of composite systems.
Therefore, all peculiarities related to the order in tensor products
should be taken into account. In particular, we should remember that
the order in a $C^*$-algebra (so in $\cB(\cH)$) is not susceptible
to the order of multiplication. The same can be said about the notion
of positive map. On the contrary the notion of completely positive map and the
notion of $m$-positive map ($m>1$) are susceptible to the order of multiplication.
This explains the role of transposition $\tau$, its relation to $M_n^+ \otimes M_k^+$,
$(M_n \otimes M_k)^+ $ (in Choi's scheme), and finally
the significance of the proper chosen family of functionals in definition
of the order in the tensor product. This 
concluding ``mathematical'' remark can be also considered as a partial answer 
to the question related to
some experiments of the Bell type, like those studied in (\cite{zuk}).
Namely, there are
problems associated with propagation properties (variables) of photons as well
as with corellations between their spins. 
But such a system is an example of a composite system.
Moreover we note that any real experiment involves time
evolution. So, a description of such a system deals with dynamical maps
preserving an order structure of tensor product. Furthermore, an
analysis of such the systems involves spin variables as well as 
other ones necessary for full description.
Consequently, in general, the considered maps should be sensitive
to the order properties of various variables. 
Taking the separability problem as an illustration we have argued that
basic features of the problem really follow from such the general theory.

\vskip 0.3cm
{\bf Acknowledgments:} 
We are grateful to M. Zukowski and M. Marciniak for remarks. 
W.A.M. acknowledges the support from K.B.N. (project 
PB/0273/PO3/99/16 ).


\begin{references}
\bibitem{buba}
We recall that to each physical system one can associate
the set $\cal O$ of all its observables, the set $\cal S$ of
all its states and a mapping $<{\cal O}, {\cal S}> \to \Rn$ which
associates with each pair $(A, \varphi)$ in $({\cal O}, {\cal S})$
a real number $\varphi(A)$ that we interpret as the expectation
value of the observable $A$ when the system is in the state
$\varphi$. Then, for mathematical convinience, the above
mentioned map can be extended over a set containing also non-hermitian
operators, e.g. over ${\cal B}(\cal H)$. Since such extended map is in one-to-one
correspondence with a state we can term, with a slight abuse of language, 
this map with the name state.
\end{references}
\end{document}